\documentstyle[12pt]{article}
\begin{document}
\def\om{\omega}
\def\omt{\tilde{\omega}}
\def\ti{\tilde}
\def\o{\Omega}
\def\t{T^*M}
\def\vt{\tilde{v}}
\def\ot{\tilde{\Omega}}
\def\otwo{\omt \wedge \om}
\def\owot{\om \wedge \omt}
\def\w{\wedge}
\def\mt{\tilde{M}}
\def\ss{\subset}
\def\tpm{T_{P} ^* M}
\def\al{\alpha}
\def\alt{\tilde{\alpha}}
\def\inop{{\int}^{P}_{P_{0}}{\om}}
\def\th{\theta}
\def\tht{\tilde{\theta}}
\def\inox{{\int}^{X}{\om}}
\def\inotx{{\int}^{X}{\omt}}
\def\st{\tilde{S}}
\def\l{\lambda}
\def\p{{\bf{p}}}
\def\pb{{\p}_{b}(t,u)}
\def\pbm{{\p}_{b}}
\def\del{\partial}
\def\delb{\bar{\partial}}
\def\l2{\Lambda^2}
\def\be{\begin{equation}}
\def\ee{\end{equation}}
\def\ej{{\bf E}}
\def\ed{{\bf E}^\perp}
\def\d{\cdot}
\def\si{\sigma}
\def\cg{{\cal G}}
\def\cgt{\ti{\cal G}}
\def\bz{\bar{z}}
\def\e{\varepsilon}
\def\b{\beta}
\begin{titlepage}
\begin{flushright}
CERN-TH/95-39\\
hep-th/9502122
\end{flushright}
\vskip 1cm
\begin{center}
{\Large\bf Dual Non-Abelian
 Duality \\  and the Drinfeld Double} \\
\vskip 1cm
{\bf C. Klim\v c\'\i k
\footnote{CERN Fellow. On leave from Charles University, Prague.}} \\
\vskip 0.3cm
  {\it Theory Division CERN, CH-1211 Geneva 23,
Switzerland} \\
\vskip 0.5cm {\small and} \\
\vskip 0.5cm
 {\bf P. \v Severa}\\
\vskip 0.3cm
 {\it Dept. of Theoretical Physics, Charles
University,} \\ {\it V Hole\v sovi\v ck\'ach 2, CZ-18000 Prague,
 Czech Republic} \\
\end{center}
\vskip 0.5cm
\begin{abstract}

The standard notion  of the non-Abelian duality in string theory
is  generalized to the class of
$\si$-models admitting `non-commutative conserved charges'. Such
$\si$-models
can be associated with every Lie bialgebra $(\cg ,\cgt)$ and they
possess an
isometry group iff the commutant  $[\cgt,\cgt]$ is not equal to
$\cgt$.
 Within the enlarged class of the backgrounds the non-Abelian duality
{\it is} a duality transformation in the proper sense of the word. It
exchanges the roles of $\cg$ and $\cgt$ and it can be interpreted as
a
symplectomorphism of the phase spaces of the mutually dual theories.
We give
explicit formulas for the non-Abelian duality transformation for any
$(\cg,\cgt)$. The non-Abelian analogue of the Abelian modular space
$O(d,d;{\bf Z})$ consists of all maximally isotropic decompositions
of the
corresponding Drinfeld double.

\end{abstract}
\vskip 0.5cm
CERN-TH/95-39\\
February 1995
\end{titlepage}

\section{\bf Introduction}
Duality symmetry plays an important role in string theory as a
tool for disentangling its full symmetry structure. It deepens our
understanding of the geometry of the spacetime from the string point
of view,
because it relates apparently different, but still equivalent,
backgrounds
in the $\si$-model approach. The Abelian target space duality kept
receiving
the attention of string theorists in the past few years and much
progress has
been achieved in revealing the consequences of the symmetry in
string theory
and in classifying backgrounds related by the Abelian duality group
$O(d,d;{\bf Z})$ \cite{Busch,Duff,RoVe,Kir,Alv1,GiRa,Alv2}. From the
$\si$-model point of view the necessary condition to work out  a dual
to some
background was that the latter possess an Abelian group of
isometries. Thus,
the class of string backgrounds concerned was rather restricted and
many
physically relevant classical string vacua were excluded from the
consideration.

An important materialization of the suspicion that the duality
symmetries
should be  associated also with the non-Abelian isometries of the
target
manifold was achieved by de la Ossa and Quevedo \cite{OQ}. They
gauged  the
non-Abelian isometries of $\si$-models and constrained the field
strenth $F$
to vanish. The dual action was then obtained by integrating out the
gauge
fields and the Lagrange multipliers had become coordinates of the
dual
manifold. In the series of subsequent investigations
\cite{GR,GRV,AAL,EGRSV,AAL2,S,ST,BS,Ty} other relevant insights were
obtained, still the notion of `non-Abelian duality' was lacking some
of the
key features of its Abelian counterpart. The non-Abelian isometry
group of
the dual space was always smaller;  also, a canonical procedure was
missing
that would yield the original theory if one is given only its
non-Abelian dual. For the above reasons it was somewhat considered as
a
misnomer to call the non-Abelian duality a duality transformation
\cite{Alv1}.

In this contribution we  attempt to cure those drawbacks of the
non-Abelian
duality. The starting point of our considerations was the fact that a
non-Abelian dual, even without isometries, is still equivalent to an
apparently different $\si$-model. This made us believe that the
relevant
algebraic structure  for the existence of a non-Abelian T-duality is
not
necessarily the existence of the group of the isometries of the
background,
but some
other structure that shows up only in special cases as an isometry
group.
We  even show that what is called the non-Abelian duality in the
present
nomenclature, could  also be referred to as a sort of semi-Abelian
duality.
Indeed, as a result of our analysis, we are able to present a lot of
examples
of the  full non-Abelian duality, when the duality is
performed in both directions without the isometry.

In looking for an explicit criterion, when a given $\si$-model is the
non-Abelian dual of another one,  the clear geometric
understanding of the Abelian duality, which we presented in our
previous
work \cite{KS}, proved very useful. In the formalism developed there
the central role was played
by the Noetherian forms on the world-sheet associated with the
Abelian
isometry group $G_0$ of the target. Those forms were closed (hence
integrable)  on any extremal string surface by  virtue of the
symmetry
 and their integrals turned out to be the coordinates of the dual
target.
Both the original $M$ and the dual target $\ti M$ were embedded into
one manifold
$M_E$ which turned out to have a natural structure  of a fibre bundle
over
$M/G_0$ (or $\ti M  / \ti G_0$) with the fibre $G_0 \times \ti G_0$.
The
canonical symplectic structure on $G_0 \times \ti G_0$ gave the
difference
between the original and the dual action, or, in other words,
Buscher's
formula \cite{Busch}.

All relevant steps of the previous construction can be repeated for a
much
general class of targets, however. We again need the action of a
(possibly
non-Abelian) group $G$ on the $\si$-model  target which gives rise to
the
Noetherian forms. If $G$ does not act as an isometry, the forms are
not
closed even on the extremal surfaces. They can still be integrable,
however!
Imagine that we organize the Noetherian forms $\al_a ~
(a=1,2,...,dimG)$ into
a Lie algebra valued form $\al=\al_a\ti T^a$, where $\ti T^a$ are the
generators of some (dual) Lie algebra $\cgt$. Suppose that the target
has the
property that on the extremal string surfaces the form $\al$ is a
flat
connection, i.e. it satisfies the Maurer-Cartan equation
   \be  d\al_a-{1\over 2}\ti c^{~kl}_a \al_k\w\al_l=0,    \ee
where $\ti c^{kl}_a$ are the structure constants of $\cgt$. Then the
form
$\al$ is integrable, which means that there exists a map $\ti
g(\tau,\sigma)$
from the worldsheet to the dual group $\ti G$ such that
   \be  \al=d\ti g~\ti g^{-1}.        \ee
We call this property a {\it non-commutative conservation law}. If
the
group $G$ acts freely on the original target then we can choose a
preferred
system of coordinates $(y,g)$ where $y$'s label the orbits of $G$ in
the
target $M$ and $g\in G$. We shall see that if the connection $\al$ is
flat,
extremal strings live naturally in an extended manifold $M_E$ having
the
structure of a fibre bundle over $M/G$ (or $\ti M / \ti G$) with the
fibre
being  the Drinfeld double $(G,\ti G)$. The canonical symplectic
structure on
the double gives the difference between the original and the dual
action and,
hence, the non-Abelian Buscher's formula. The group $\ti G$ acts
naturally on
the dual target $\ti M$ and the dual Noetherian form can be organized
in the
$\cg$-valued form $\ti\al=\ti\al_a T^a$. So the procedure can be
repeated,
returning to the original target.

The analogue of the Abelian modular space $O(d,d,{\bf Z})$ is given
by the
structure of the Drinfeld double, namely by the classification of the
decompositions of the algebra of the double in the pairs of maximally
isotropic
subalgebras with respect to the $ad$-invariant bilinear form on the
double.
Such decompositions can be constructed by means of the automorphisms
of the
Drinfeld double, which naturally form a subgroup of
$O(dim\cg ,dim\cgt ,{\bf Z})$. However, unlike  in the Abelian case,
the
automorphisms
 do not necessarily exhaust all possibilities.

The standard non-Abelian duality of de la Ossa and Quevedo \cite{OQ}
is the
special case of our treatment. The dual group $\ti G$ is Abelian and
the
corresponding Drinfeld double is the cotangent bundle  of the group
manifold
$G$ with its canonical symplectic form. The Abelian duality is
described by
the double where both groups are Abelian. It has the topology of a
$2dimG$-dimensional torus and its group of automorphisms (preserving
the
invariant bilinear form) is $O(dimG,dim\ti G,{\bf Z})$.

In section 2 of our note we give the explicit criterion when a given
$\si$-model is the non-Abelian dual of another  and a straightforward
prescription of how to reconstruct the original model from its dual
(or how to
perform the
non-Abelian duality in both directions). We emphasize that no
relevant local
algebraic structure is lost upon performing the duality
transformation. We
describe in detail the geometric structure of the non-Abelian duality
by
means of the lift of the dynamical characteristics of string to the
Drinfeld
double.

In  section 3 we shall discuss the interpretation of the
non-Abelian duality in terms of a canonical transformation. In
section 4
we give explicit formulas for the non-Abelian duality transformations
for a
generic Lie bialgebra and discuss their projective character.

In the concluding section 5 we  describe  the relevant structure
of the Drinfeld double which gives rise to the non-Abelian analogue
of the
Abelian modular space $O(d,d,{\bf Z})$; we discuss the dressing
transformations and
touch the issue of integrability. We finish with comments about
quantization,
in particular about the conformal invariance, the dilaton and
possible
emergence of the quantum group structure.

\section{\bf Non-Abelian duality and Lie bialgebras}

In what follows we shall consider two-dimensional $\si$-models
described
by a metric $G_{ij}$ on the target manifold $M$ and a globally
defined two-form
$B_{ij}$ on $M$ with the action
\be S=\int dz d\bz (G_{ij}(x)+B_{ij}(x))\del x^i \delb x^j\equiv\int
dz d\bz
E_{ij}\del x^i \delb x^j .\ee
Suppose that a group $G$ acts freely on $M$. We can associate to this
action
the Noetherian forms on the world-sheet given by
\be J_a=v_a^i(x)E_{ij}\delb x^j d\bz-v_a^i(x)E_{ji}\del x^j dz,\ee
where $v_a^i(x)$ are the (left-invariant) vector fields corresponding
to the
right action of $G$ on $M$. They can be defined also when $G$ is not
the
isometry of the target, by varying the action with respect to the $G$
transformations with the world-sheet dependent parameters
$\e^a(z,\bz)$, i.e.
\be \delta S=S(x+\e^a v_a)-S(x)=\int \e^a{\cal{L}}_{v_a}(L) +
\int d\e^a\w J_a.\ee
If the Lie derivative of the Lagrangian ${\cal{L}}_{v_a}(L)$
vanishes, then
the forms $J_a$ are closed on the extremal surfaces $x^i(z,\bz)$. We
shall
look for a condition on $E_{ij}$ which would guarantee that the forms
$J_a$
on the extremal surfaces satisfy
\be dJ_a={1\over 2}\ti c_a^{~kl}J_k\w J_l.\ee
Here $\ti c_a^{~kl}$ are the structure constants of some Lie algebra
$\cgt$.
 From Eq. (3) it follows that
\be {\cal{L}}_{v_a}(L)={1\over 2}\ti c_a^{~kl}J_k\w J_l\ee
or, in other words,
\be {\cal{L}}_{v_a}(E_{ji})=\ti c_a^{~kl} v_k^m v_l^n
E_{mi}E_{jn}.\ee
If the condition (8) holds we may associate to each extremal surface
$x^i(z,\bz)$ a mapping $\ti g(z,\bz)$ from the world sheet into the
dual
group $\ti G$ such that
\be J_a=d\ti g~\ti g^{-1}\ee
or
\be \ti g=P\exp{\int_{\gamma}J_a\ti T^a},\ee
where $P$ means the path-ordered exponential.
We shall refer to the Lagrangians fulfilling (8) as to the
$\si$-models
admitting
non-commutative conservation laws.

 Note from (10) that $\ti g$ is defined
up to the homotopy class of the curve $\gamma$. If, for instance, we
integrate around a closed non-contractible loop on the world sheet of
the
closed string, the integral (10) gives a fixed element of the dual
group
$\ti G$, which we refer to
as a `charge'. However,  this charge is not a number but a
non-commutative
object. If we run around the loop twice we have to multiply charges
rather
than add them.

Condition (8)  in fact requires that a certain compatibility
requirement
should be imposed on the structure constants of the original and dual
Lie
algebras. This requirement is the integrability condition of the set
of the
first-order differential equations (8). It is easy to see that the
condition
reads
\be \ti c_k^{~ac}c^l_{~fa}-\ti c_k^{~al}c^c_{~fa}-\ti c_f^{~ac}c^l_{~ka}
+\ti  c_f^{~al}c^c_{~ka}-\ti
c_a^{~lc}c^a_{~fk}=0.\ee
Amazingly, this is the standard relation which must be obeyed by the
structure
constants of the Lie bialgebra $(\cg,\cgt)$ \cite{D,AM,FG}! This
condition is
manifestly dual, hence we expect that there exists an equivalent dual
$\si$-model where the roles of $\cg$ and $\cgt$ are exchanged.
Obviously, the
dual model $\ti E_{ij}$ should fulfil\be {\cal{L}}_{\ti v_a}(\ti
E_{ij})=
c_a^{~kl}\ti v_k^m \ti v_l^n \ti E_{mi}\ti E_{jn}.\ee
For the sake of clarity we shall first discuss the case in which the
group
$G$ acts on the target transitively (and freely), i.e. the target
itself can
be identified with the group manifold. Then there is a very easy and
beautiful way of
solving Eqs. (8) and (12), using the concept of the Drinfeld double
$D$ \cite{D,AM,FG}. The latter is the connected group  corresponding
to the
Lie algebra double
$\cal{D}$ and containing both groups $G$ and $\ti G$. The double
$\cal{D}$
is
equal to $\cg +\cg^*$ as the vector space with the Lie
bracket\footnote{$\cgt$ is identified with the dual space $\cg^*$ of
$\cg$.}
\be [X+v,Y+w]\equiv [X,Y]+[v,w]^*-ad_X^* w+ad_Y^* v +ad_w^*
X-ad_v^*Y.\ee
Here $ad_X^*$ is the usual $ad^*$-operator for the Lie algebra $\cg$
acting
on $\cg^*$. The symbol $ad_w^*$ corresponds to the coadjoint action
of the Lie
algebra $\cg^*$ on its dual space $\cg$. Note that both groups $G$
and $\ti G$
are embedded into $D$. The algebras $\cg$ and $\ti \cg$ form the
maximally
isotropic subspaces of $\cal{D}$ with respect to the $ad$-invariant
non-degenerate bilinear form
\be(X+v,Y+w)\equiv \langle X,w\rangle+\langle Y,v\rangle.\ee
The symbol $\langle .,.\rangle$ means the canonical pairing between
the algebra $\cg$ and
its dual $\cg^*$.

Consider the tangent space $T_e D\cong {\cal{D}}$ at the unit element
$e\in{\cal{D}}$. (Of course, $e$ is the unit of both $G$ and $\ti G$
at the
same time.) In $T_e D$ we can take a $d$-dimensional subspace
$\cal{E}$ which
is the graph\footnote{By the graph we mean the set
$\{t\in\cg, t+E(t,.)\}\ss\cg+\cgt$.}
of a non-degenerate linear mapping $E:\cg\to\cgt$. The subspace
${\cal{E}}\ss T_e D$ can be transferred to every point $
g\in G(\hookrightarrow D)$ by the right action of $G$ itself. At the
point
$g\in G$ $T_g D\cong{\cal{D}}$ again and its decomposition into $\cg
+\cgt$
is given by the {\it left} action of $g$ on $T_e D$.
Hence, we have defined at every $g\in G$ a non-degenerate linear
mapping
$E_g:\cg\to\cgt$ with the graph ${\cal{E}}_g$. Since $\cgt$ is
canonically
identified with $\cg^*$, we have obtained a matrix $E_{ab}(g)$. It is
straightforward to check that $E_{ab}(g)$ solves Eq. (8)\footnote{In
fact, it
is the general solution, $\cal{E}$ playing the role of the initial
value for
the first-order equation (8).}. Obviously, the solution of Eq. (12)
can be
obtained in the same way by transferring a subspace $\ti {\cal{E}}\ss
T_e D$
into the whole $\ti G$ by the right action of $\ti G$ itself. It is
natural
 (and
also supported
by the Abelian duality case \cite{KS}) to conjecture that the
mutually dual
$\si$-models are obtained by taking $\ti {\cal{E}}={\cal{E}}$. In
other
words, we transfer the same subspace ${\cal{E}}\ss T_e D$ onto $G$
and $\ti G$.

Now we have to prove that the $\si$-models $E_g$ and $\ti E_{\ti g}$
are
equivalent. First of all we map every solution $g(z,\bz)$ of the
original
model into a solution  $\ti h(z,\bz)$ of the dual one. Following
 Eq. (9), $g(z,\bz)$ can be considered as a surface
\be f(z,\bz)=g(z,\bz)\ti g(z,\bz)\ee
in the Drinfeld double $D$, where the multiplication is taken in $D$.
It is
known \cite{AM} that the following two decompositions are applicable
for every
$f\in D$:
\be  f(z,\bz)=g(z,\bz)\ti g(z,\bz)= \ti h(z,\bz) h(z,\bz).\ee
We show that $\ti h(z,\bz)\in \ti G$ defined by Eq. (16) is indeed a
solution
of the dual model $\ti E_{\ti g}$ and $h(z,\bz)$ is associated to it
by the
dual analogue of Eq. (9). The easiest way to show this consists in
finding
a condition when a surface $l(z,\bz)$ in $D$ can be obtained by
lifting an
extremal solution of the model $E_g$ via (15). The conditions read
\be \delb l~l^{-1}\in{\cal{E}},\qquad \del
l~l^{-1}\in{\cal{E}}^{\perp},\ee
where the orthogonal complement is taken with respect to the bilinear
form
(14). Before proving the statement, note that the conditions (17) do
the right
job because they do not depend in any way on the choice of the group
$G$ or
$\ti G$ in the double $D$. Thus the existence of the extremal $\ti
h(z,\bz)$
with
the associated $h(z,\bz)$ is obvious, since only the extremal
surfaces are
liftable. The choice $\ti {\cal{E}}={\cal{E}}$ is crucial for the
statement,
of course.

The proof that the conditions  of `liftability'
are given by the relations (17) requires a little geometry. Suppose
that
an element $g\in G(\hookrightarrow D)$ lies on the surface $l(z,\bz)$
in $D$.
A vector $\del g=\e^a v_a$ at $g$ on an extremal surface in $G$ is
lifted
into $T_g D$ via Eq. (9), i.e. it becomes
\be \del f=\e^a v_a +\e^a J_a=\e^a v_a-E(.,\e^a v_a).\ee
The last equality follows from the definition (4) of the currents. In
a
similar way
\be \delb f=\delb g+E(\delb g,.).\ee
Clearly,  if $l(z,\bz)$ is the
liftable surface, $\del l$ and $\delb l$ have to obey Eqs.
(18) and (19). In
other words
\be \delb l\in{\cal{E}}_l, \qquad \del l\in{\cal{E}}_l^{\perp}.\ee
Because ${\cal{E}}_l$ was obtained from ${\cal{E}}$ by the right
action of $l$,
the conditions (17) follow. If we are at a point $l(z,\bz)$ which
does not
lie at {\it G} we may transfer it there by the right action of $\ti
G$, because the
lift of the extremal surface in $G$ via Eq. (9) is defined up to the
right
action of $\ti G$.
\section{\bf Non-Abelian duality as a symplectomorphism}

So far we have constructed the mapping between the phase spaces of
the
original and the dual $\si$-models. This mapping has a constructive
character
because it
guarantees that solving the original $\si$-model we can also solve
the
apparently different $\si$-model. In this sense the two theories are
equivalent.
We shall
show, however, that the equivalence of the models can be understood
in a much
stronger way, namely the mapping between the phase spaces preserves
their
natural symplectic structure \footnote{For the special case of the
non-Abelian
duality between the $SU(2)$ group and its coalgebra the statement has
been
proved in \cite{CZ}.}. To demonstrate this we have to extend our so
far local
analysis by the discussion of the boundary conditions. According to
the
comments
after Eq. (10), the integration of the original Maurer-Cartan form
(4) around
the non-contractible loop on the world sheet of the closed string
gives the
non-commutative charge belonging to the dual group. We shall restrict
the
phase spaces of the models to the configurations having the unit
charge,
otherwise the lifting of the string into the Drinfeld double does not
give a
closed loop. Such a restriction renders degenerate the natural
symplectic form
$\Omega_{Ph}$ coming from the action. To cure the problem we have to
perform
a generalized Marsden-Weinstein reduction. As  already mentioned, the
lift $g(z,\bz)\ti g(z,\bz)$ of an extremal surface into the double is
defined
up to the right multiplication by a constant element $\ti g_0\in\ti
G$. All
such lifts we identify in the dual phase space and we proceed
similarly in
the dual case. Only applying this procedure does the mapping between
(the reduced)
phase spaces become one-to-one and,
moreover, the restricted form $\Omega_{Ph}$ becomes non-degenerate.
Now we
prove that the non-Abelian duality is a symplectomorphism of the
(reduced)
phase spaces.

 Let $LG$ be the loop space of the target $G$. As usual,
 we obtain the phase space from the cotangent bundle $T^*LG$, on
 which there is the canonical symplectic form $\o _{Ph} =d \th
_{Ph}$.
 Namely, we identify some submanifold in $T^*LG$ and then factorize
 it appropriately\footnote{We proceed conceptually as in the case
 of a relativistic particle in a background; in the $\sigma$-model
 case the submanifold is defined by the Virasoro constraints.}.
 The construction goes as follows: if we have a string world-sheet
$F$
 and a loop $l$ on it, then we define a corresponding element
 $l_F \in T^*_l LG$. To describe how $l_F$ acts on a vector
 $u \in T_l LG$, first realize that $u$ can be thought of as a
 family of vectors $u(X) \in T_X G$ where $X$ runs along $l$. Then
\be l_F (u) \equiv \oint _l J_a u^a(X). \ee
If we take all $l_F$'s for all possible $F$'s we obtain the mentioned
 submanifold of $T^*LG$. Now we identify all $l_F$'s coming from the
 same extremal $F$ and obtain the (unreduced) phase space.

 Let $H$ be a
surface (i.e. a 2-parametric family of
on-shell strings) in the unreduced phase space of the original model
such
that all strings have the unit charge. Let there also be a loop on
each
$F \in H$
$l(F)$.
 Then by (21)
\be\int _H \o _{Ph} = \oint _{\del H} \th _{Ph}
 = \oint\limits_{\bigcup_{F \in \del H} l(F)} J \ee
Here $J$ is understood as a two-form on the two-dimensional closed
surface
$\bigcup_{F \in \del H} l(F)$ in the target $G$. The form $J_a$ on
the
world-sheet is to be saturated by the vectors on the surface tangent
to the
loops $l(F)$ and the index $a$ is saturated by the vectors connecting
the
infinitesimally close loops $l(F)$.
Now we lift the family $H$ into a family $H_D$ of surfaces in $D$;
$H_D$ is
defined up to an independent shift by the right action by a constant
element
from $\ti G$ on each surface in $H_D$. We project the family $H_D$
into
$\ti G$ according Eq. (16), thus obtaining the family $\ti H$ of
extremal
surfaces in $\ti G$. We have to  prove that
\be
 \oint\limits_{\bigcup_{F \in \del H} l(F)}J =
 \oint\limits_{\bigcup_{\ti F \in \del \ti H} l(\ti F )}\ti J .\ee
We stress that this relation holds in spite of the ambiguity in the
definition of $\ti H$. This means that the symplectic form
$\ti\Omega_{Ph}$
is well defined on the {\it reduced} dual phase space. The dual
statement
holds, too.

We shall compare the two expressions in Eq. (23), using the common
lifted family
 $H_D$. We demonstrate that if $t_D$ and $u_D$ are vectors
 at a point $P$ of a lifted surface $F^D$, $t_D$ tangent to $F^D$ and
 $u_D$ arbitrary, then
\be\ti J_a(\ti t)\ti u^a-J_a(t)u^a=\Omega_D(t_D \w u_D),\ee
where $\ti t,\ti u\in T\ti G$; $t,u\in TG$ are given by the
projections
(inverse lifts) and $\Omega_D$ is the canonical symplectic form on
$D$
\cite{AM}, to be
written explicitly in what follows. Consider the subspaces $S_{R(L)}$
and
$\ti S_{R(L)}$ obtained by the right (left) action of $P$ on
$\cg+\cgt$
embedded in $T_e D$. We now define a linear mapping $\Pi_{R\ti R}$ in
$T_P D$
as the projection on $\ti S_R$ with the kernel $S_R$ and accordingly
for
the other combinations of the indices. By definition, $\ti J_a(\ti
t)\ti u^a=
(t_D,\Pi_{L\ti R}u_D)$ and $J_a(t)u^a=(t_D,\Pi_{\ti L R}u_D)$, where
the
round bracket is the invariant bilinear form (14) in the double.
According to
Ref. \cite{AM}\footnote{We are much indebted to B. Jur\v co for
pointing out
to  us the existence of the  symplectic structure $\Omega_D$, which
we badly
needed.}
\be \Omega_D(t_D,u_D)=(t_D,(\Pi_{R\ti R}-\Pi_{\ti L L})^{-1}u_D).\ee
Thus we have to prove that
\be (\Pi_{R\ti R}-\Pi_{\ti L L})(\Pi_{L\ti R}-\Pi_{\ti L R})=1.\ee
We do it easily
\begin{eqnarray}  &(\Pi_{R\ti R}-\Pi_{\ti L L})(\Pi_{L\ti R}-\Pi_{\ti
L R})=
\cr &=\Pi_{R\ti R}\Pi_{L\ti R}-\Pi_{\ti L L}\Pi_{L \ti R}+\Pi_{\ti L
L}
\Pi_{\ti L R}
=\Pi_{L\ti R}-\Pi_{\ti L L}\Pi_{L\ti R}+\Pi_{\ti L L}=\cr
&=(\Pi_{L\ti R}+
\Pi_{\ti R L})+(\Pi_{\ti L L}-\Pi_{\ti R L}-\Pi_{\ti L L}\Pi_{L\ti
R})=1+0.
\end{eqnarray}
Now from Eq. (24) we can conclude
\be \oint\limits_{\bigcup_{\ti F \in \del \ti H} l(\ti F )} \ti J-
 \oint\limits_{\bigcup_{F \in \del H} l(F)} J=
\oint\limits_{\bigcup_{F^* \in \del H^* }l(F^*)}\Omega_D =0\ee
because $\Omega_D$ is closed and the closed surface over which we
 integrate is a boundary.

\section{\bf  Projective transformations}
In this section we give explicit formulas for the non-abelian duality
transformations for a generic Lie bialgebra $(\cg,\cgt)$. They easily
follow
from the general discussion in section 2. When the group $G$ acts
transitively on the target, the explicit formula for the $\si$-model
$E_g$ is
given by
\be E_g^t=d(g)E_0^t(a(g)+b(g)E_0^t)^{-1},\ee
where $t$ means the transposition of matrices and the functions
$a,b,d$ are
the components of the adjoint action of $g$ on
${\cal{D}}=\cg+\cgt$. In other words
\be g^{-1}\left(\matrix{X\cr v}\right)g\equiv
\left(\matrix{a(g)&b(g)\cr 0&d(g)}\right)\left(\matrix{X\cr
v}\right),\ee
where $X\in\cg$ and $v\in\cgt$. The dual $\si$-model is obtained by
\be \ti E_{\ti g}^t=\ti d(\ti g){E_0^t}^{-1}(\ti a(\ti g)+
\ti b(\ti g){E_0^t}^{-1})^{-1}.\ee
Note that the constant matrix $E_0^t$ in (29) was replaced by its
inverse in
(31) because the subspace $\cal{E}$ in $T_e D$ is the graph of the
inverse of
$E_0^t$ from the dual point of view.

We may illustrate the content of  formulas (29) and (31) for the case
of
the
non-abelian duality of de la Ossa and Quevedo \cite{OQ}. The double
$D$ is
simply the cotangent bundle $T^*G$ with the structure of the
semi-direct
product of $G$ and the abelian group ${\bf R}^{dimG}$. For simplicity
we take
$E_0=Id$. Then Eq. (29) gives
\be E_g=Id\ee
and Eq. (31)
\be (\ti E_{\chi}^{-1})^{ab}=\delta^{ab}+\chi^k c_k^{ab},\ee
where $\chi^k$ are coordinates on the fibre. The corresponding
Lagrangians
are respectively
\be L=Tr(g^{-1}\del g g^{-1}\delb g)\ee
and
\be \ti L=\ti E_{ab}(\chi)\del \chi^a \delb \chi^b. \ee

 We now present the analogues of the Buscher formula for the abelian
duality.
We consider the case in which $G$ does not act transitively. The
coordinates
labelling the orbits of $G$ in the target $M$, we denote
$y^{\al}(\al=1,\dots,n)$. The matrix of the $\si$-model $E_{ij}$ has
both
types of indices corresponding to $y^{\al}$ and $g$. The Lagrangian
reads
\begin{eqnarray} &L=E_{\al\b}(y)\del y^{\al}\delb y^{\b}+
E_{\al b}(y,g)\del y^{\al}(g^{-1}\delb g)^b+\cr &+
E_{a\b}(y,g)(g^{-1}\del g)^a\delb y^{\b} +
E_{ab}(y,g)(g^{-1}\del g)^a (g^{-1}\delb g)^b.\end{eqnarray}
Note that the dependence of $E_{ij}$ on $g$ is fixed by  condition
(8).
Explicitly
\be E^t(y,g)=D(g)E^t(y,e)(A(g)+B(g)E^t(y,e))^{-1},\ee
where $e$ is the unit element of $G$, $E(y,e)$ can be chosen
arbitrarily and
$A(g)$ is the $(n+dimG)\times(n+dimG)$ matrix
\be A(g)\equiv\left(\matrix{Id &0\cr 0&a(g)}\right),\qquad B(g)\equiv
\left(\matrix{0 &0\cr 0&b(g)}\right)
\ee
and $D(g)$ is given in terms of $d(g)$ in the same way as $A(g)$ in
terms of
$a(g)$. Needless to say, $a(g),b(g)$ and $d(g)$ are the same as in
Eq. (29).
As far as the dual model $\ti E$ is concerned
\be \ti E^t(y,\ti g)=\ti D(\ti g)\ti E^t(y,e)(\ti A(\ti g)+
\ti B(\ti g)\ti E^t(y,e))^{-1}.\ee
Here
\be \ti E^t(y,e)=(C+DE^t(y,e))(A+BE^t(y,e))^{-1},\ee
where
\be A=D=\left(\matrix{Id&0\cr0&0}\right),\qquad B=C=
\left(\matrix{0&0\cr0&Id}\right).\ee
\section{Non-abelian modular space and conclusions}
Even without a clear Abelian motivation, the natural question to ask
is what is the modular space $\cal{M}$ of the $\si$-models equivalent
by the $(\cg,\cgt)$ duality. By  equivalence we mean that solving one
$\si$-model in the modular space $\cal{M}$,  the solutions of all
other models in $\cal{M}$
follow. It certainly does not come as a surprise that $\cal{M}$ is
given by
the structure of the Drinfeld double $D$. Suppose that $\cal{D}$ can
be
decomposed differently, say in $({\cal{K}},\ti {\cal{K}})$, in such a
way
that both algebras ${\cal{K}}$ and $\ti {\cal{K}}$ are maximally
isotropic
subspaces of $\cal{D}$ with respect to the $ad$-invariant bilinear
form (14).
A $({\cal{K}},\ti{\cal{K}})~ \si$-model can be obtained by the right
action
of the group $K$ on the subspace ${\cal{E}}\ss T_e D$ (see section
2). If the
subspace $\cal{E}$ is the same as the corresponding subspace defining
the
$(\cg,\cgt)$ model then two models  are necessarily equivalent.
Indeed,
condition (17) for a surface $l(z,\bz)\in D$ to be obtained by
lifting an
extremal solution is the same whether the lifting is done from $G$ or
from $K$, depending just on the subspace
${\cal{E}}\in T_e D$. The explicit form of the solution $k(z,\bz)$
associated
to a solution $g(z,\bz)$ is found by using the decomposition
(16) from the point of view of $({\cal{K}},\ti{\cal{K}})$, i.e.
\be g(z,\bz)\ti g(z,\bz)=k(z,\bz)\ti k(z,\bz).\ee
It seems  an interesting problem to find all maximally isotropic
decompositions for a generic double $D$. We did not attempt to do
that, but we
should remark that the modular space $\cal{M}$ is not, in general,
exhausted
just by the automorphisms of the double\footnote{It is so in the
purely
abelian case where the modular space $O(d,d,{\bf Z})$ is just the
group of
the automorphisms of the abelian double.}. Indeed, the pure
$Z_2$-duality
$(\cg,\cgt)\to(\cgt,\cg)$ is not an automorphism of the double if the
algebras $\cg$ and $\cgt$ have a different structure.

Although we can relate two theories in the modular space $\cal{M}$
without
actually solving them, it would be interesting to know whether the
rich
algebraic structure underlying the models can help to do that.
Answering this
question is somewhat indirectly related to the issues discussed in
this note;
nevertheless, we should probably mention how the lifting of the
extremal
surfaces into the double $D$  naturally leads to a sort of the
dressing
transformation
\cite{ZS,DJKM,STS,BB}
generating new solutions from a known one. The key point is to
realize that
the condition (17), when a surface $l(z,\bz)$ in $D$ is the lift of
some
extremal solution from $G$ (or $\ti G$), is invariant with respect to
the
right multiplication of $l(z,\bz)$ by an arbitrary constant element
of $D$.
Suppose that $g(z,\bz)$ is an extremal surface in $G$. Its lift into
$D$ is
given by $g(z,\bz)\ti g(z,\bz)$. Now $g(z,\bz)\ti g(z,\bz) g_0$ is
also the
lift of some extremal surface from $G$, i.e.
\be g(z,\bz)\ti g(z,\bz) g_0=g_1(z,\bz)\ti g_1(z,\bz),\ee
where $g_1(z,\bz)$ is determined from $g(z,\bz),\ti g(z,\bz)$ and the
constant element $g_0\in G$. In other words, although the group $G$
does not
act  on the target $G$ as the isometry of the $\si$-model,  its
action on the
double $D$ via Eq. (43) does yield new solutions of the model from a
known one.

So far our discussion was purely classical. It is obviously of utmost
interest whether the described non-Abelian dualities relate conformal
field
theories (CFT) or can be even interpreted as exact symmetries of the
CFT.
Some interesting results were obtained in \cite{S}, where it was
shown that
some gauged WZNW models based on the non-semi-simple algebras are
equivalent
to the non-Abelian duality transformations of the WZNW actions.
It is tempting to conjecture that  some gauged $G/H$ WZNW models
could
possess the bialgebra structure. Since extracting the classical
geometry of
the target is somewhat involved procedure \cite{T} it may be
difficult to see
immediately whether they admit the non-commutative conservation laws.
After
all, a real understanding of the  sense in which the non-Abelian
duality is
the
symmetry of the CFT requires  carrying out in detail the operator
mapping
between a given theory and its non-Abelian dual. As a prerequisite
for such
investigations, we need the derivation of the $(\cg,\cgt)$
non-Abelian
duality by a sort of path integral manipulations. Although we have
made some
progress in this direction, which we do not present here, this is
still not
sufficient to yield the complete solution of the problem. It is
certainly one
of the most important open issues which have to be settled in order
to
proceed further with the CFT application, with the problem of dilaton
which
we have completely ignored and with a description of the Abelian
duality
resembling that given by  Ro\v cek and Verlinde
 \cite{RoVe}. Another interesting problem
would consist in understanding the non-Abelian duality when the
groups do
not act freely. Let us conclude in an optimistic way: we believe that
the
rich algebraic structure of the presented models may turn out to be
sufficient for performing a consistent quantization of the theories
involved,
hopefully yielding also some non-trivial CFT. It is clear that in
that case we
may expect new and deep applications of quantum groups in string
theory.
\section{\bf Acknowledgement}
We thank L. \'Alvarez-Gaum\'e, B. Jur\v co and E. Kiritsis for
discussions.


\begin{thebibliography}{19}
\bibitem{Busch}{T.H. Buscher, Phys. Lett. B194 (1987) 51, B201 (1988)
466.}
\bibitem{Duff}{M.J. Duff, Nucl. Phys. B335 (1990) 610}
\bibitem{RoVe}{M. Ro\v cek and E. Verlinde, Nucl. Phys. B373 (1992)
630}
\bibitem{Kir}{A. Giveon and M. Ro\v cek, Nucl. Phys. B380 (1992) 128;
A. Giveon and E. Kiritsis, Nucl. Phys. B411 (1994) 487; E. Kiritsis,
Nucl. Phys. B405 (1993) 109}

\bibitem{Alv1}{E. \'Alvarez, L. \'Alvarez-Gaum\'e, J. Barb\'on and
 Y. Lozano, Nucl. Phys. B415 (1994) 71.}
\bibitem{GiRa}{A. Giveon, E. Rabinovici and G. Veneziano,
 Nucl. Phys. B322 (1989) 167; K.A. Meissner and G. Veneziano,
 Phys. Lett. B267 (1991) 33}
\bibitem{Alv2}{E. \'Alvarez, L. \'Alvarez-Gaum\'e and Y. Lozano,
Phys. Lett. B336 (1994) 183}
\bibitem{OQ}{X. de la Ossa and F. Quevedo, Nucl. Phys. B403 (1993)
377}
\bibitem{GR}{A. Giveon and M. Ro\v cek, Nucl. Phys. B421 (1994) 173}
\bibitem{GRV}{M. Gasperini, R. Ricci and G. Veneziano, Phys. Lett.
B319 (1993) 438}
\bibitem{AAL}{E. \'Alvarez, L. \'Alvarez-Gaum\'e and Y. Lozano, Nucl.
Phys.
B424 (1994) 155}
\bibitem{EGRSV}{S. Elitzur, A. Giveon, E. Rabinovici, A. Schwimmer
and
G. Veneziano, {\it Remarks on Non-abelian Duality}, CERN-TH-7414/94,
hep-th/9409011}
\bibitem{AAL2}{E. \'Alvarez, L. \'Alvarez-Gaum\'e and Y. Lozano,
{\it An Introduction to T-Duality in String Theory}, CERN-TH-7486/94,
hep-th/9410237}
\bibitem{S}{K. Sfetsos, Phys. Rev. D50 (1994) 2784}
\bibitem{ST}{K. Sfetsos and A.A. Tseytlin, Nucl. Phys. B427 (1994)
245}
\bibitem{BS}{I. Bakas and K. Sfetsos, {\it T-Duality and World-Sheet
Supersymmetry}, CERN-TH-95-16, hep-th/9502065}
\bibitem{Ty}{E. Tyurin, {\it On Conformal Properties of the Dualized
$\si$-models}, ITP-SB-94-58, hep-th/9411242}
\bibitem{KS}{C. Klim\v c\'\i k and P. \v Severa, {\it Strings in
Space-Time
Cotangent Bundle and T-duality}, CERN-TH-7490/94, hep-th/9411003, to
appear
in Mod. Phys. Lett. A}
\bibitem{D}{V.G. Drinfeld, {\it Quantum Groups}, in Proc. ICM, MSRI,
Berkeley,  1986, p. 708}
\bibitem{AM}{A.Yu. Alekseev and A.Z. Malkin, Commun. Math. Phys. 162
(1994)
147}
\bibitem{FG}{F. Falceto and K. Gaw\c{e}dzki, J. Geom. Phys. 11 (1993)
251}
\bibitem{CZ}{T. Curtright and C. Zachos, Phys. Rev. D49 (1994) 5408}
\bibitem{ZS}{V.E. Zakharov and A.B. Shabat, Funct. Anal. Appl. 13
(1979) 166}
\bibitem{DJKM}{E. Date, M. Jimbo, M. Kashiwara and T. Miwa, Physica
4D (1982)
343}
\bibitem{STS}{M.A. Semenov-Tian-Shansky, Publ. RIMS, Kyoto Univ. 21
(1985)
1237}
\bibitem{BB}{O. Babelon and D. Bernard, Commun. Math. Phys. 149
(1992) 279}
\bibitem{T}{A.A. Tseytlin, Nucl. Phys. B399 (1993) 601}
\end{thebibliography}
\end{document}